\documentclass{JHEP3}

\usepackage{graphicx}
\usepackage{dcolumn}
\usepackage{bm}
\usepackage{slashbox}
\usepackage{epsfig}

\newcommand{\be}{\begin{equation}}
\newcommand{\ee}{\end{equation}}
\newcommand{\bea}{\begin{equationarray}}
\newcommand{\eda}{\end{equationarray}}

\newcommand{\rnc}{\renewcommand}

\rnc{\baselinestretch}{1.24}    
\rnc{\arraystretch}{1.24}   

\makeatletter \rnc{\theequation}{\thesection.\arabic{equation}}
\@addtoreset{equation}{section} \makeatother

\def\be{\begin{equation}}
\def\ee{\end{equation}}
\def\ba{\begin{array}}
\def\ea{\end{array}}
\def\bea{\begin{eqnarray}}
\def\eea{\end{eqnarray}}
\def\nn{\nonumber\\}

\def\ct#1{\cite{#1}}

\def\eq#1{Eq. (\ref{#1})}

\def\g{\gamma}

\def\k{\kappa}

\def\L{\Lambda}

\def\th{\theta}

\def\r{\rho}

\def\half{\frac{1}{2}}

\def\ll{\left\langle}
\def\rr{\right\rangle}

\def\lc{\left\{}
\def\ls{\left(}

\def\rc{\right\}}
\def\rs{\right)}

\def\Tr{{\rm Tr}\,}

\title{Holographic trace anomaly at finite temperature}

\author{Bum-Hoon Lee\\ Department of Physics, Sogang University, Seoul, Korea 121-742 and\\ CQUeST, Sogang University, Seoul, Korea 121-742\\ E-mail: \email{bhl@sogang.ac.kr}}

\author{Siyoung Nam\\ CQUeST, Sogang University, Seoul, Korea 121-742\\ E-mail: \email{stringphy@gmail.com}}

\author{Chanyong Park\\ CQUeST, Sogang University, Seoul, Korea 121-742\\ E-mail: \email{cyong21@sogang.ac.kr}}

\abstract{
We find an exact coordinate transformation rule from the $AdS_5$ Schwarzschild
black hole in the Poincare and the global patch to the Fefferman-Graham
coordinate system. Using these results, we evaluate the corresponding
holographic stress tensor and trace anomaly of the boundary theory as a
function of the radial coordinate. Following the AdS/CFT correspondence,
we reinterpret the radial coordinate dependence of the trace anomaly
as the Wilsonian renormalization group(RG) flow of the boundary theory.
}

\maketitle

\newpage

\begin{document}

\tableofcontents


\section{Introduction}
The idea of the AdS/CFT correspondence \cite{Maldacena:1997re, Gubser:1998bc,
Witten:1998qj}(For reviews, see \cite{Aharony:1999ti, D'Hoker:2002aw}) is that string
theory or M-theory in the near-horizon geometry of a collection of coincident D-branes
or M-branes is equivalent to the low-energy world-volume theory of the corresponding branes.
For example, the AdS/CFT correspondence relates type IIB superstring theory or M-theory
in space-time geometries that are asymptotically anti-de Sitter(AAdS) times a compact space
to conformal
field theories(CFTs).
The conjectured AdS/CFT correspondence can be regarded as a realization of the
holographic principle \cite{'tHooft:1993gx, Susskind:1994vu} and offer an example of
a gravity/gauge theory duality. The strong coupling regime of the quantum field
theory(QFT) defined on a background geometry which is conformally related to the
geometry at the boundary of the AAdS space corresponds to the weak coupling regime
of the string theory and vice versa. This implies that, at least in principle, we
can obtain information on one side of the duality by performing computations on the
other side.

According to the AdS/CFT correspondence, for every bulk field there
is a corresponding gauge invariant boundary operator. The explicit prescription
of \cite{Gubser:1998bc, Witten:1998qj} is equating a partition function of the
classical gravity theory which is the low energy limit of the string theory
\begin{equation}
Z_{SUGRA}[\phi_0 ] = \int_{\phi_0} {\cal{D}}\phi \exp
\left(-S [\phi, g_{\mu \nu} ] \right)
\end{equation}
to a generating functional of the dual CFT
\begin{equation}
Z_{CFT}[\phi_0] = \left< \exp \int_{boundary}d^dx \phi_0 {\cal{O}} \right>
\end{equation}
where $\phi_0$ denotes the boundary value of the field $\phi$.
For example, the boundary value of the gravitational field is
the source of the stress tensor of the dual theory. However,
the boundary of the $AdS$ spacetime locates at the spatial infinity and the
Weyl factor will diverge at the boundary due to the infinite volume factor
of the on-shell action. Thus, we must regularize the theory and introduce
counter terms to properly remove the infinities. The counter terms which are
local functionals of the intrinsic geometry of the boundary are chosen to diverge
at the boundary in such a way as to cancel the bulk divergences. After that,
we can construct the finite boundary stress tensor. This procedure is called
as a `counter term' subtraction method\cite{Balasubramanian:1999re, deHaro:2000xn,
Henningson:1998ey}. We can do this in the Poincare patch or the global patch of the
five-dimensional Schwarzschild-AdS(SAdS) black hole.

On the other hand, we can also do this in the so-called Fefferman-Graham(FG) coordinates.
According to \cite{FG}, the AAdS metric near the boundary allows a perturbative
expansion\footnote{$i,j= 0,1,2,3$ stands for the boundary coordinates and
$\mu,\nu = 0,1,2,3,4$ for the bulk coordinates. We also denote the spatial
coordinates by a, b=1,2,3.}
\begin{eqnarray} \label{fgz1}
ds^2 ~=~ G_{\mu\nu}dx^{\mu}dx^{\nu} ~=~ \frac{l^2}{z^2}dz^2
+ \frac{l^2}{z^2}g_{ij}(x,z) dx^idx^j
\end{eqnarray}
where
\begin{equation} \label{fgz2}
g_{ij}(x,z)=g^{(0)}_{ij}+g^{(2)}_{ij}z^2+g^{(4)}_{ij}z^4+\cdots
+g^{(d)}_{ij}z^d+h^{(d)}_{ij}z^d\ln z^2+\cdots\, .
\end{equation}
This is the solution of the Einstein equation in the Fefferman-Graham
coordinate system.
Originally, Fefferman and Graham considered a solution of
AAdS space-time with $\textbf{S}^d$ boundary for even $d$, where the logarithmic
term appears naturally.
As will be shown, in the cases of SAdS black hole with the boundary topology
$\textbf{R} \times \textbf{R}^3$ (in the Poincare patch) and
$\textbf{R}\times \textbf{S}^3$ (in the global patch), there is no logarithmic
term in (\ref{fgz2}) so  we will drop the logarithmic term from now on.
Actually, the Euclidean versions of the above boundaries are
$\textbf{S} \times \textbf{S}^3$ and $\textbf{S}\times \textbf{R}^3$ in the global
patch and the Poincare patch, respectively.


In this paper, we will evaluate the boundary stress tensor from the bulk
gravity side using AdS/CFT and study the radial dependence of the stress
tensor by adopting a Wilsonian renormaization group(RG) perspective. This
task will be undertaken by using the exact black hole solution in the FG
coordinates for the Poicare and the global patch. From the point of view of
the metric (\ref{fgz1}), the boundary conditions are imposed at some finite
cut-off value $z=z_{0}$, not at $z=0$ (which would be a true boundary
of AdS)\cite{Klebanov:2000me}. Since the effective energy scale on the
boundary theory is associated with the radial coordinate of the bulk
space-time \cite{Maldacena:1997re}, the physical meaning of the
cut-off $z_0$ is that it acts as a UV regulator in the boundary
theory\cite{Gubser:1998bc, Susskind:1998dq}. The
bulk diffeomorphism that induces a Weyl transformation on the boundary
metric is reminiscent of renormalization group(RG)
flow\cite{deHaro:2000xn, deBoer:1999xf, Imbimbo:1999bj}. So when the boundary
is located at a finite distance in the radial direction, the position of the
boundary can be reinterpreted as an energy cut-off of the boundary theory in
the Wilsonian RG sense, where the energy `cut-off' means that we have to
integrate out
all higher energy modes than this cut-off scale.

For this Wilsonian RG interpretation of the boundary theory, we need
a well-defined boundary metric up to overall scale corresponding to
the conformal structure of the boundary theory\cite{deHaro:2000xn,
deBoer:1999xf, Imbimbo:1999bj}.
In the usual $r$-coordinate, this conformal structure of the boundary theory
is not clear but in the FG coordinate this is manifest.
So although we know a metric in the $r$-coordinate,
to investigate the structure of the boundary theory it is important to find
the same solution in the FG coordinate. After finding this solution in the FG
coordinate, we have to choose the boundary metric as $g_{ij}(x,z)$ to
obtain the finite boundary stress tensor
\be
T_{ij} \equiv \frac{2}{\sqrt{-g}}\frac{\delta S}{\delta g^{ij}} .
\ee
When considering the boundary metric
as $\g_{ij}(x,z)  =\frac{l^2}{z^2}g_{ij}(x,z)$ \ct{Balasubramanian:1999re},
the boundary stress
tensor diverges when $z \to 0$. From now on, we consider $g_{ij}(x,z)$
as a boundary metric.

In section 2 we give a brief review of the (minimal) counter term subtraction
method. In section 3 we find an exact black hole solution in the FG
coordinate and calculate the boundary energy-momentum tensor as a function
of the bulk radial coordinate. Using these results, we can investigate the
trace anomaly depending on the radial position of the boundary, which is
reinterpreted as a Wilsonian RG cut-off. In section 4 we finish this paper
with some discussion and conclusion.


\section{Constructing Finite Boundary Stress Tensor}
In this section, we briefly review the two methods how to construct the finite
stress tensor of the dual boundary theory. One of them is the (minimal) counter term
subtraction method and the other is the holographic renormalization.

First we start by discussing the counter term subtraction method.
One of the interesting problems
in general relativity is how to construct a local stress tensor of a given
theory in the curved background.  In \cite{Brown:1992br},
Brown and York(BY) suggested a quasi-local stress tensor, which is defined as the
functional derivative of the on-shell action with respect to the boundary metric
and introduced the reference frame to render the action finite. This is called as
a `background' subtraction method. According to the AdS/CFT correspondence,
the quasi-local stress tensor corresponds to the expectation value of the stress
tensor in the dual CFT. Balasubramanian and Kraus(BK) constructed the Brown-York
tensor for asymptotically AdS space time with a minimal set of the possible
counter term\cite{Balasubramanian:1999re}. For convenience, we call this counter
term substraction method as BK method .

Let us consider the five-dimensional space time $\cal{M}$ foliated by four-dimensional
time like hyper surfaces $M_z$ homeomorphic to the boundary $\partial\cal{M}$. We choose
such a coordinate system that $x^i$ denotes the coordinates tangential to the
hyper surfaces and $z$ the normal coordinate.
The total $AdS_5$ action with a minimal counter term action is
\begin{eqnarray} \label{EHaction}
S&=&\frac{1}{2\kappa^2}\int_{\cal{M}}d^5 x\sqrt{-G}(\textrm{R}+\frac{12}{l^2})
-\frac{1}{\kappa^2}\int_{\cal{\partial M}} d^4 x\sqrt{-\gamma}K\nonumber\\
&& \quad -\frac{3}{l\kappa^2}\int_{\cal{\partial M}} d^4 x\sqrt{-\gamma}
\left( 1-\frac{l^2}{12}R^{(4)}\right)
\end{eqnarray}
where $\kappa^2=8\pi G$ and $G$ is the five-dimensional gravitational constant
and $R^{(4)}$ is the curvature scalar constructed from the boundary metric
$\gamma_{ij}$. The first term in the action is the ordinary Einstein-Hilbert
action with a cosmological constant $-6/l^2$ where $l$ is the curvature radius
of the anti-de Sitter space. The second term is the Gibbons-Hawking boundary
term defined by the trace of the extrinsic curvature
\begin{equation}
K_{ij}~\equiv~ -\nabla_{i}n_{j}
\end{equation}
which is needed for a well-defined variation of the action.
The last term is the counter term which makes the action finite
at the boundary.
Then, the quasi-local stress tensor (or Brown-York tensor) $<T^{BY}_{ij}>$ is given by
\begin{eqnarray}   \label{BK}
<T^{BY}_{ij}>&=&\frac{2}{\sqrt{-\gamma}}\frac{\delta S}{\delta\gamma^{ij}}\nonumber\\
&=&\frac{1}{\kappa^2}\left(K_{ij}-K\gamma_{ij} -\frac{3}{l}\gamma_{ij}
+\frac{l}{2}G^{(4)}_{ij}\right)\, ,
\end{eqnarray}
where $G^{(4)}_{ij} \equiv R^{(4)}_{ij}-(1/2)R^{(4)}\gamma_{ij}$ is
an Einstein tensor of the boundary theory. Let us consider the following AAdS metric
\be
ds^2 = \frac{l^2}{z^2}\left( g_{ij} dx^i dx^j + dz^2\right).
\ee
When the spatial boundary of the AAdS space time locates at $z=0$, the
boundary metric $\g_{ij} = (l^2/z^2) g_{ij} $ has
a double pole. So to obtain a well-defined stress tensor
we have to redefine the boundary metric
by an appropriate scale transformation as $g_{ij} = (z^2/l^2) \g_{ij} $.
Then, the well-defined boundary energy-momentum tensor
is given by
\be     \label{bst}
T_{ij} \equiv \frac{2}{\sqrt{-g}}\frac{\delta S}{\delta g^{ij}}
= \frac{l^2}{z^2} T^{BY}_{ij}\, ,
\ee
which gives a finite value at the boundary of the AAdS space time \cite{Myers:1999ps}.

Next, we assume that the boundary is at some finite value $z=z_0$.
From the UV-IR relation, the boundary position $z_0$ can be reinterpreted
as an energy scale  of the boundary theory $\L = 1/z_0$.
From the viewpoint of Wilsonian RG sense, this implies that
the higher energy modes of the boundary theory are integrated out.
Therefore, this geometric setting describes the physics of the
boundary theory at the given energy scale $\L$. Since inserting
a scale can usually break the conformal symmetry,
the conformal symmetry of the boundary theory at $z_0$ is also
broken so that we expect that there exists a trace anomaly depending
on $\L$. As will be shown, the effect of this energy scale is encoded
to the finite size effect in the global patch, where the boundary
has a $\textbf{R} \times \textbf{S}^3$ topology. When considering the black hole, there is
another scale corresponding to the black hole mass, which also break the
conformal symmetry and appears as a thermal effect of the boundary theory.

We outline another method, `holographic renormalization',
to evaluate the holographic stress tensor in the FG
coordinates\cite{deHaro:2000xn,Henningson:1998ey}.
According to the AdS/CFT correspondence, the generating
function of a conformal field theory is given in terms
of the on-shell gravitational action on ${\cal{\partial
M}}$, which is proportional to the volume
Vol(${\cal{\partial M}}$). The volume
Vol(${\cal{\partial M}}$) of any conformally compact
manifold $\cal{M}$ is infinite. So an appropriate
renormalization of Vol(${\cal{\partial M}}$) must be
carried out. This procedure starts from regulating the theory. To do this,
we restrict the range of the radial coordinate to $\rho \ge \r_0$ and think
the boundary as located at $\rho=\r_0$.\footnote{Here, we use the variable
$\rho=z^2$ for the sake of the convenient calculation, because the expansion
(\ref{fgz2}) contains only even powers of z.}
Therefore, the regulated action is given by
\begin{eqnarray}
S_{reg}&=&\frac{1}{2\kappa^2}\int_{\rho\ge \r_0}d^5x
\sqrt{-G}(\textrm{R}-2\Lambda) -\frac{1}{\kappa^2}\int_{\rho=\r_0}d^4x
\sqrt{-\gamma}K\nonumber\\
&=&\frac{l^3}{2\kappa^2}\int d^4x \sqrt{-g_{(0)}} \left[\frac{-6}{\rho_0^2}
+\frac{1}{2}\left(\Tr[g^{-1}_{(0)}g_{(2)}g^{-1}_{(0)}g_{(2)}] \right. \right. \nn
&& \left. \left. \quad \quad \quad \quad
-\Tr[g^{-1}_{(0)}g_{(2)}]\Tr[g^{-1}_{(0)}g_{(2)}]\right)
\ln\rho_0+{\cal{O}}(\rho_0^2)\right]\, .
\end{eqnarray}
After subtracting the divergences by the counter terms and
removing the regulator, we obtain the holographically renormalized action
\begin{eqnarray}
S_{ren}&=& \lim_{\r_0\rightarrow 0}\frac{l^3}{2\kappa^2}\left[S_{reg}
-\int d^4x \sqrt{-g_{(0)}} \lc \frac{-6}{\rho_0^2}
+\frac{1}{2}\left(\Tr[g^{-1}_{(0)}g_{(2)}g^{-1}_{(0)}g_{(2)}]
\right. \right. \right. \nn
&& \left. \left. \left. \qquad \qquad -\Tr[g^{-1}_{(0)}g_{(2)}]
\Tr[g^{-1}_{(0)}g_{(2)}]\right)\ln\rho_0\rc \right]
\end{eqnarray}
Then, the holographic stress tensor can be obtained by the following
definition\cite{deHaro:2000xn}
\begin{equation}
<T^{0}_{ij}>=\frac{2}{\sqrt{-g_{(0)}}}\frac{\partial S_{ren}}{\partial g^{ij}_{(0)}}\, ,
\end{equation}
which becomes in terms of $g_{(0)}, ~ g_{(2)}$ and $g_{(4)}\, $
\begin{eqnarray}    \label{sd}
<T^{0}_{ij}>&=&\frac{l^3}{\kappa^2}\left[2g^{(4)}_{ij}-\frac{1}{4}
\left[\Tr[g^{-1}_{(0)}g_{(2)}]\Tr[g^{-1}_{(0)}g_{(2)}]
- \Tr[g^{-1}_{(0)}g_{(2)}g^{-1}_{(0)}g_{(2)}]\right]g^{(0)}_{ij}\right.\nonumber\\
&&\left.-\left(g_{(2)}g^{-1}_{(0)}g_{(2)}\right)_{ij}
+\frac{1}{2}\Tr[g^{-1}_{(0)}g_{(2)}]g^{(2)}_{ij}\right]\, .
\end{eqnarray}

As will be shown with an explicit calculation in the next section,
this holographic stress tensor corresponds to
the leading term of the boundary stress tensor defined in (\ref{bst}),
\be
<T^{0}_{ij}>=\lim_{\r_0 \rightarrow 0} T_{ij}\, ,
\ee
which implies that at $\r_0 (z_0) \to 0$ the higher order corrections like
$g_{(6)}, ~ g_{(8)}, \cdots$ are suppressed.

\section{Trace Anomaly in Schwarzschild-AdS Black Hole}

\subsection{Anomaly in the Poincare Patch}

In the Poincare patch, a black hole (or black brane) metric is given by
\begin{equation}
ds^2 = - f(r) dt^2 + \frac{dr^2}{f(r)} + \frac{r^2}{l^2}d\vec{x}^2\,  ,
\end{equation}
with $f(r) = r^2/l^2 -m/r^2$. In the above equation,
\begin{equation} \label{pbhmass}
m \equiv \frac{2\kappa^2M l^3}{3V_{3}}
\end{equation}
is related to a black hole mass $M$
and a volume of the three dimensional space $V_3$.
The black hole horizon $r_h=(ml^2)^{1/4}$
satisfies $f(r_h)=0$ and the black hole entropy proportional to the area at this
horizon is given by
\be \label{ep}
S = \frac{m^{3/4} V_3 }{4G l^{3/2}}\, .
\ee
Using the surface gravity $K \equiv(1/2) \partial_r f(r) |_{r=r_h}$,
the Hawking temperature is given by
\be \label{tp}
T_H \equiv \frac{K}{2 \pi}= \frac{m^{1/4}}{\pi l^{3/2}}\,  .
\ee
Note that when $m=0$ this black hole metric reduces to a metric for a pure
AdS space-time.

To obtain a black hole solution in the FG coordinate system, we must evaluate
the following integral
\be  \label{intg}
\int\frac{dr}{\sqrt{f(r)}} ~=~ -\int \frac{l dz}{z}\, ,
\ee
the (d+1)-dimensional result of which is given by
\begin{equation}
r(z) ~=~ l\left[ \left(\frac{l}{z}\right)^{\frac{d}{2}} + \frac{m}{4l^{d-2}}
\left(\frac{z}{l}\right)^{\frac{d}{2}}\right]^{\frac{2}{d}}\, .
\end{equation}
In this paper, we are interested in the five-dimensional case ($d=4$) and
thus the relation becomes
\begin{equation} \label{ctrans}
r(z) = \frac{l^2}{z}\sqrt{1+\frac{m z^4}{4l^6}}\, .
\end{equation}
Then, the black hole metric in the FG coordinates\cite{Nakamura:2006ih} is
\begin{equation}
ds^2 = -F(z) dt^2 +G(z) d\vec{x}^2 + \frac{l^2}{z^2}dz^2
\end{equation}
where
\begin{equation}
F(z) ~=~ \frac{l^2}{z^2}\frac{\left(1-\frac{m z^4}{4l^6} \right)^2}{(1
+\frac{m z^4}{4l^6})} ~~~\textrm{and}~~~
G(z) ~=~ \frac{l^2}{z^2} \left(1+\frac{m z^4}{4l^6}\right)\, .
\end{equation}
From the coordinate transformation (\ref{ctrans}), we can easily find
the relations among the functions $F(z)$, $G(z)$, the lapse function $f(r)$,
and the coordinate $r$ as a function of $z$ :
\begin{equation}
F(z) ~=~ f(r(z)) ~~~\textrm{and}~~~
G(z) ~=~ \frac{r(z)^2}{l^2}\, .
\end{equation}
The equation $F(z) = 0$ gives a radius of the event horizon
$z_h = \sqrt{2} l^{3/2} / m^{1/4}$
, which is related to $r_h$ under the coordinate transformation.
The entropy proportional to the area at the event horizon is given by
\begin{equation}    \label{ef}
S =\frac{V_3}{4G} \sqrt{G(z_h)^3} = \frac{ m^{3/4} V_3}{4G l^{3/2}}\,  .
\end{equation}
The Hawking temperature in the FG coordinate is given by
\begin{equation} \label{htempz}
T_H = \left.\frac{1}{4 \pi} \partial_r f(r)\right|_{r=r_h} = \left.\frac{1}{4\pi l}
\frac{\partial_z F(z)}{\partial_z \sqrt{G(z)}}\right|_{z=z_h}\,  ,
\end{equation}
where $l \partial_z \sqrt{G(z)}$ is due to the coordinate
transformation. From this formula, the Hawking temperature in the FG coordinate becomes
\begin{equation}    \label{tf}
T_H = \frac{m^{1/4}}{\pi l^{3/2}}\, ,
\end{equation}
which gives the same Hawking temperature, as expected.
In the FG coordinate, we can easily obtain the thermodynamic quantities of
the black hole by using (\ref{ef}) and (\ref{tf}), which gives the exactly
same results as those in the usual $r$-coordinate, (\ref{ep}) and (\ref{tp}).

Now, we evaluate the boundary stress tensor and the trace of it. First we
follow the BK-method using (\ref{bst}). Here, only the following counter
term is needed
\begin{equation}
S_{ct} = -\frac{1}{\kappa^2}\int d^4x\sqrt{-\gamma}~\frac{3}{l}\, ,
\end{equation}
because the Einstein tensor of the boundary metric $\gamma_{ij}$ is trivially
zero in the Poincare patch. Then, the boundary stress tensor is given by
\bea \label{pbst}
\kappa^2 \ll T_{00} \rr
&~=~&\frac{3m}{2 l^3}\cdot\left(\frac{4 l^6 - m \r^2} {4 l^6 + m \r^2 }\right)^2
\texttt{}\, , \nn
\kappa^2 \ll T_{aa} \rr
&~=~&\frac{m}{2 l^3}\cdot \left(\frac{4 l^6 +  3 m \r^2 }{4 l^6 - m \r^2}\right) .
\eea
This is an exact result describing the boundary stress tensor at
the RG scale $\L = 1/\sqrt{\r}$ and at the finite temperature $T_H = \frac{m^{1/4}}{\pi l^{3/2}}$.
The trace of the stress tensor is given by
\be \label{tap}
\k^2 \ll T^i_{\phantom{a}i} \rr ~=~ \frac{24 m^2 l \r^3}{(4l^6 - m \r^2) (4l^6+m\r^2)}\, .
\ee

Next, we calculate the stress tensor by the holographic renormalization method.
Using the formula (\ref{sd}) we get the result
\begin{equation}
\kappa^2<T^{0}_{ij}> ~=~ \frac{m}{2l^3}\, \textrm{diag}(3,~1,~1,~1)\, .
\end{equation}
For comparison, we expand the exact results (\ref{pbst}) and (\ref{tap})
near $\r = 0$
\begin{eqnarray} \label{epbst}
\kappa^2 \ll T_{00} \rr &=&\frac{3m}{2l^3}-\frac{3m^2 \rho^2}{2 l^{9}}
+\frac{3m^3\rho^4}{4 l^{15}}+{\cal{O}}(\rho^6)\, ,\nonumber\\
\kappa^2 \ll T_{aa} \rr &=&\frac{m}{2 l^3}+\frac{m \rho^2}{2 l^{9}}
+\frac{m^3 \rho^4}{8 l^{15}}+{\cal{O}}(\rho^6)\, ,\nonumber\\
\k^2 \ll T^i_{\phantom{a}i} \rr &=&  \frac{3 m^2 \r^3}{2 l^{11}}
+ \frac{3 m^4 \r^7}{32 l^{23}} + {\cal O}(\r^{11})\, .
\end{eqnarray}
Then, as expected, the leading part of the stress tensor $<T_{ij}>$ defined by the
BK-method corresponds to the stress tensor $<T^{0}_{ij}>$ acquired by the
holographic renormalization. These leading terms are also calculated in \cite{Myers:1999ps}.
Furthermore, we can also write the stress tensor in the compact form.
Using the equations (\ref{tf}) and (\ref{tap}), the stress tensor is rewritten by
\begin{equation}
<T_{ij}> ~=~ <T^i_{\phantom{a}i}> C(\rho) g_{ij}
\end{equation}
where
\begin{equation}
C(\rho) ~=~ \frac{l^2}{4\rho}+\frac{l^2(1 - 4 \delta_{i0}
\delta_{j0})}{3\pi^4 T_H^4 \rho^3}\,  .
\end{equation}
It is easy to check that, in the limit $\rho\rightarrow 0$,
$<T^i_{\phantom{a}i}> \sim \rho^3$, $C(\rho)\sim 1/\rho^3$ and
$g_{ij}\sim \rho^0$, so that the leading term of the stress tensor is
independent of the radial coordinate $\rho$. Note that the stress tensor
is proportional to the boundary metric.
By following the method of \cite{Balasubramanian:1999re}, we can also calculate the mass
\begin{equation}
M ~=~\lim_{\rho\rightarrow 0}\frac{3mV_3}{8\kappa^2l^9}|m\rho^2-4l^6|\, ,
\end{equation}
which confirms the normalization of the black hole mass density
(\ref{pbhmass}) and agrees with the result in \cite{Balasubramanian:1999re,
 Horowitz:1998ha}. According to the AdS/CFT correspondence, this stress tensor can be
reinterpreted as that of the boundary theory. From the viewpoint of the
boundary theory, the black hole mass $m$ corresponds to the temperature
of the boundary theory with (\ref{tf}) and the radial coordinate $\r$ is
reinterpreted as an inverse energy scale $\L = 1 / \sqrt{\r}$ of the boundary
theory. Hence we can write the trace in terms of the boundary variables,
$\L$ and $T_H$
\begin{eqnarray}
\kappa^2 \, T^i_{\phantom{a}i}\, &=& \,
\frac{24\sqrt{\Lambda}}{\frac{16}{\pi^8}\left(\frac{\Lambda}{T_H^4}\right)^4-1} \, \nonumber \\
&=& \, \frac{3\pi^8 T_H^8}{2\Lambda^{3/2}} \left[1+
\frac{\pi^8}{16}\left(\frac{T_H^4}{\Lambda}\right)^2
+\frac{\pi^{16}}{256}\left(\frac{T_H^4}{\Lambda}\right)^4 +\cdots \right]
\end{eqnarray}
Since the radial coordinate dependence of the quantities in the bulk theory
can be considered as a Wilsonian RG flow of the boundary theory, $\L$ is
introduced as an energy scale of the Wilsonian RG flow in the boundary theory
where the higher energy modes than $\L$ are integrated out. At the UV fixed
point $\r = 0$ where the trace anomaly is zero, the boundary stress tensor
satisfy the Stefan-Boltzman law\cite{Nakamura:2006ih, Gubser:1996de}
\be
\kappa^2 \ll \textrm{T}_{00} \rr ~=~ \frac{3 \pi^4 l^3}{2}~T_H^4\, .
\ee
When $m=0$, this SAdS black hole solution reduces to the pure AdS space-time
solution with a boundary $\textbf{R} \times \textbf{R}^3$.
As is well known, the gravity theory on the pure
AdS space-time corresponds to the N=4 SYM which is superconformal. Therefore,
we can easily expect that there is no trace anomaly for N=4 boundary SYM.
In (\ref{tap}), for $m \to 0$ the trace of the boundary theory becomes zero
which implies that the boundary theory is always conformal theory independent
of the scale. For $m \ne 0$, this describes the finite temperature field theory
defined on the Hawking temperature $T_H = m^{1/4}/(\pi l^{3/2})$. At the
UV fixed point $\r=0$, the trace anomaly of this boundary theory is zero so
this theory is still conformal. When $\r \ne 0 $, the trace anomaly is not
zero due to the thermal correction. This trace anomaly grows monotonically as
$\r$ goes to the black hole horizon $\r_h = z_h$ and at the black hole horizon
this trace anomaly diverges (see figure 1).

\begin{figure} \label{ptra}
\begin{center}
\includegraphics[height=.3\textheight, width=.5\textwidth]{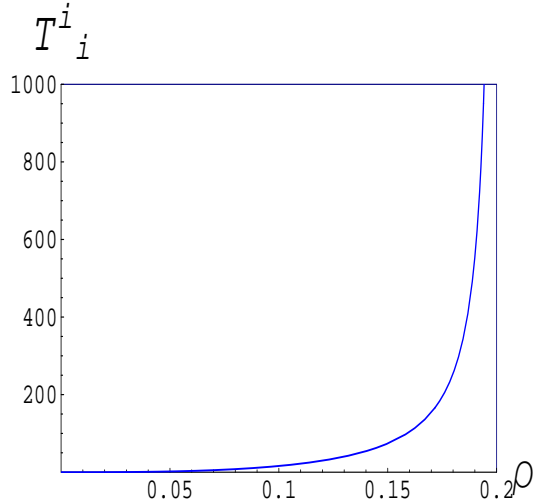}
\caption{Trace Anomaly in the Poncare patch.  For convenience, we set $l=1$
and $\kappa^2=1$. With $m=100$, the trace anomaly diverges at the black hole
horizon $\rho_h=0.2$.}
\end{center}
\end{figure}

\subsection{Anomaly in the Global Patch}


A SAdS black hole solution in the global coordinate is given by the following expression
\begin{equation}   \label{bhm}
ds^2 ~=~ - f(r)dt^2 + \frac{1}{f(r)} dr^2 + r^2 d \Omega_3^2 \, ,
\end{equation}
where
\begin{equation}
f(r)~=~ 1 +\frac{r^2}{l^2} - \frac{m}{r^2}
\end{equation}
and $d\Omega_3^2~=~w_{ab}dx^adx^b~=~d\theta^2+sin^2\theta d\phi^2+cos^2\theta d\psi^2$
is a metric for a unit three-sphere $S^3$. Here, $m$ is defined by
\begin{equation} \label{gbhmass}
m=\frac{2\kappa^2M}{3V_{S^3}}
\end{equation}
where $M$ is a black hole mass and $V_{S^3}=2\pi^2$ is the volume of
the unit three sphere.  From the equation $f(r_h)=0$, the black hole
horizon $r_h$ is given by
\begin{equation}     \label{rh}
r_h = l\left(\frac{\sqrt{1+\frac{4m}{l^2}}-1}{2}\right)^{1/2}\, .
\end{equation}
Then, the Hawking temperature is given by
\begin{equation}     \label{htemp}
T_H ~=~ \frac{1}{2\pi l}\cdot
\sqrt{\frac{2+8m/l^2}{\sqrt{1+4m/l^2} - 1}}\,  .
\end{equation}
and the black hole entropy is
\begin{equation}     \label{ent}
S ~=~ \frac{\pi^2}{2G}r_h^3 ~=~ \frac{\pi^2 l^3}{2G}
\left(\frac{\sqrt{1+\frac{4m}{l^2}}-1}{2}\right)^{3/2}\,  .
\end{equation}
To know the metric in the FG coordinate,
we have to know the coordinate transformation between $r$ and $z$ by evaluating
the integral (\ref{intg}), which is difficult to perform.

So instead of solving (\ref{intg}), we will find the metric in
the FG coordinate by solving the Einstein equation with an appropriate metric
ansatz. We make the following metric ansatz
\begin{equation}     \label{ansatz}
ds^2 = - A(z) dt^2 + B(z) d \Omega_3^2 + \frac{l^2}{z^2} dz^2\,  .
\end{equation}
Although this metric solution describes the same SAdS black hole in the usual
$r$-coordinate system, it is a very important solution for the application of the
AdS/CFT correspondence and the exact form is not known yet.
To solve the Einstein equations exactly, we need two boundary conditions,
one of which is that this solution is given by the pure AdS space-time
near $z=0$. The other boundary condition will be fixed later.
Using the ansatz (\ref{ansatz}) and the above boundary condition,
we find an exact solution for the Einstein equation
\begin{equation}     \label{nsol}
ds^2 =  l^2\left[- \frac{\left[(1-\frac{z^2}{4l^2})^2 + cz^4 +
\frac{z^2}{2l^2} -2\right]^2}{z^2 \left[(1-\frac{z^2}{4l^2})^2+ cz^4 \right]} dt^2
+  \frac{dz^2}{z^2} +
\frac{ l^2 \left[ (1-\frac{z^2}{4l^2})^2 + cz^4\right]}{z^2} d \Omega_3^2\right]\,   ,
\end{equation}
where $c$ is an arbitrary constant which appears because of not imposing the
other boundary condition. Since our solution is another representation
of SAdS black hole solution in FG coordinate, $c$ must be identified
by $m/(4l^6)$ comparing with (\ref{bhm}) which is the same as imposing
the other boundary condition. Comparing two black hole solutions (\ref{bhm})
and (\ref{nsol}), we find that these two solutions are related by a coordinate transformation
\begin{equation} \label{transf}
r(z) = \frac{l^2}{z}\sqrt{\left(1-\frac{z^2}{4l^2}\right)^2 + \frac{mz^4}{4l^6} }\, .
\end{equation}
It is easy to check that the above equation (\ref{transf}) satisfies the
coordinate transformation integral (\ref{intg}). Using this coordinate transformation,
the functions, $A(z)$ and $B(z)$ are related to the lapse function in (\ref{bhm})
and the coordinate transformation $r(z)$ :
\begin{equation}
A(z) ~=~ f(r(z)) ~~\textrm{and}~~~ B(z) ~=~ r(z)^2\,  .
\end{equation}
In this coordinate, the black hole horizon is given by
\begin{equation} \label{rhz}
z_h =  \frac{2l}{\left(1+4m/l^2 \right)^{1/4}}\, ,
\end{equation}
which satisfy the equation $A(z_h) = 0$. The black hole entropy is given by
\begin{equation} \label{entz}
S =\frac{V_{S^3}}{4G} \sqrt{B(z_h)^3} =
\frac{\pi^2 l^3}{2G}
\left(\frac{\sqrt{1+\frac{4m}{l^2}}-1}{2}\right)^{3/2}\,  .
\end{equation}
The Hawking temperature in the FG coordinate is given by
\begin{equation} \label{htempz2}
T_H = \left.\frac{1}{4\pi}
\frac{\partial_z A(z)}{\partial_z \sqrt{B(z)}}\right|_{z_h}
= \frac{1}{2\pi l}\cdot
\sqrt{\frac{2+8m/l^2}{\sqrt{1+4m/l^2} - 1}}\, .
\end{equation}
Using these formula for the FG coordinate,
we can easily calculate the thermodynamic properties of
the SAdS black hole which are exactly the same as those obtained in the
usual $r$-coordinate.

Now, we turn our attention to evaluating the boundary stress tensor, which can be
done conveniently in the following metric, transformed by $\rho=z^2$,
\begin{equation}    \label{fgg}
ds^2 ~=~ -A(\rho)dt^2+\frac{l^2 d\rho^2}{4\rho^2}+B(\rho)d\Omega_3^2
\end{equation}
where
\begin{eqnarray}
A(\rho)&~=~& \frac{l^2}{\r}
\frac{\left[-2+\frac{\rho}{2l^2}+\frac{m\rho^2}{4l^6}+
\left(1-\frac{\rho}{4l^2}\right)^2\right]^2}{\left[\frac{m\rho^2}{4l^6}+
\left(1-\frac{\rho}{4l^2}\right)^2\right]}\, ,\nonumber\\
B(\rho)&~=~&\frac{l^4}{\r} \left[\frac{m\rho^2}{4l^6}+
\left(1-\frac{\rho}{4l^2}\right)^2\right]\, .
\end{eqnarray}
Then, it is straightforward to read off each term in the FG expansion of the
metric (\ref{fgg}) and we specify some leading terms here :
\bea \label{dss}
g^{(0)}_{ij} &~=~& {\rm diag} (-1, l^2, l^2 \sin^2 \th, l^2 \cos^2 \th )\, , \nn
g^{(2)}_{ij} &~=~& {\rm diag} (-\frac{1}{2 l^2}, -\half, -\half \sin^2 \th,
 -\half \cos^2 \th )\, , \nn
g^{(4)}_{ij} &~=~& {\rm diag} (- \frac{l^2 - 12 m}{16 l^6},
\frac{l^2 + 4 m}{16 l^4}, \frac{l^2 + 4 m}{16 l^4} \sin^2 \th,
\frac{l^2 + 4 m}{16 l^4} \cos^2 \th )\, ,  \nn
&& \vdots  ~~
\eea
Note that the perturbative expansion of (\ref{fgg}) has no logarithmic term.

Now, we evaluate the exact boundary stress tensor at finite temperature,
which is described by the black hole mass $m$ in the bulk,
with the RG scale $\L$ related to the inverse of the boundary position $1/\sqrt{\r}$.
From (\ref{bst}),
the boundary stress tensor is given by
\begin{eqnarray}
\kappa^2 \ll T_{00} \rr
&~=~&\frac{3(l^2+4m)}{8l^3}
\cdot\left(\frac{-16l^6+(l^2+4m)\rho^2} {16l^6-8l^4\rho
+ (l^2+4m)\rho^2 }\right)^2 \,  , \nonumber\\
\kappa^2 \ll T_{aa} \rr
&~=~&\frac{(l^2+4m)}{8l}\cdot
\left(\frac{16l^6-16l^4\rho+3(l^2+4m)\rho^2}{16l^6-(l^2+4m)\rho^2}\right)
\cdot w_{aa}\, .
\end{eqnarray}
Note that this is the exact boundary energy momentum tensor at arbitrary boundary
position $\r =\r_0$. Especially, at the UV fixed point ($\r \to 0$)
the boundary energy $E=V_{S^3} \ll T_{00} \rr$ and the pressure
$P_a=V_{S^3} \ll T_{aa} /w_{aa}\rr$ describes boundary fields like massless gauge
and scalar fields as a
perfect fluid type matter having the equation of state parameter $\g = P/E=1/3$.
When the boundary position is located at a finite distance $\r_0$ which
is equivalent to consider the boundary theory at the RG scale $\L$, in this scale
the interaction of the boundary fields can deform the equation of state
parameter which may depend on the RG scale $\L$ and the temperature $T=T_H$.
Using the above the exact boundary energy-momentum tensor,
we can obtain the equation of state for the boundary field at
arbitrary temperature and the RG scale, which can be written by, in terms of
the bulk variables,
\be
\g(\r,m) = \frac{(16l^6-16l^4\rho+3(l^2+4m)\rho^2)~(16l^6-8l^4\rho
+ (l^2+4m)\rho^2 )^2}{3 ~(16l^6-(l^2+4m)\rho^2)^3} .
\ee

The trace of this stress tensor is given by
\begin{equation}
\kappa^2 <T^i_{\phantom{i}i}> ~=~
-\frac{24l(l^2+4m)(4l^4-(l^2+4m)\rho)\rho^2}{(16 l^6 - (l^2 + 4 m) \rho^2)
(16 l^6 - 8 l^4 \rho + (l^2 + 4 m) \rho^2)}\, .
\end{equation}
In this global case, we can also write the stress tensor in a compact way which
is proportional to the boundary metric :
\begin{equation}
<T_{ij}>~=~C(\rho) <T^i_{\phantom{i}i}> g_{ij}
\end{equation}
where
\begin{equation}
C(\rho) ~=~ \frac{l^2}{4\rho} - \frac{l^2(\rho-4l^2)(1- 4\delta_{i0}\delta_{j0})}
{6\rho^2\left[\pi^2T_H^2 (l^2\pi^2T_H^2 +l\pi  T_H \sqrt{ l\pi T_H -2})\rho -2\right]}\, .
\end{equation}
The AdM mass is given by
\begin{equation}
M ~=~ \lim_{\rho\rightarrow 0} \frac{3\pi^2(l^2+4m)}{64\kappa^2 l^6}
|4m\rho^2+l^2\rho^2-16l^6| ~=~ \frac{3\pi^2l^2}{4\kappa^2} +\frac{3m\pi^2}{\kappa^2}\, .
\end{equation}
The second term of the mass $M$ is the standard mass of SAdS black hole solution
and the first term is the gravitational mass of global $AdS_5$ which is identified
as the Casimir energy of SU(N) ${\cal{N}}=4$ SYM on $\textbf{R}\times \textbf{S}^3$
in the large N limit\cite{Balasubramanian:1999re, Horowitz:1998ha}.

Using the
holographic renormalization method, we can easily calculate the boundary
stress tensor at the UV fixed point by inserting the
metric (\ref{dss}) into (\ref{sd})
\be \label{lbs}
\k^2 \ll T^{0}_{ij} \rr = {\rm diag} \ls
 \frac{3 (l^2 + 4m)}{8 l^3 }\,  , \frac{(l^2+4m)}{8l}\, ,
 \frac{(l^2+4m)}{8l} \sin^2 \th\, , \frac{(l^2+4m)}{8l} \cos^2 \th \rs\, .
\ee
To compare the exact result with the above boundary stress tensor,
we expand the exact one near $\r=0$ and then the several terms of
this are given by
\begin{eqnarray} \label{gtexp}
\kappa^2 \ll T_{00} \rr
&~=~&\frac{3(l^2+4m)}{8l^3}+\frac{3(l^2+4m)\rho}{8l^5}
+\frac{3(l^2-2m)(l^2+4m)\rho^2}{16l^{9}}+{\cal{O}}(\rho^3)\, ,\nonumber\\
\kappa^2 \ll T_{aa} \rr
&~=~&\left[\frac{(l^2+4m)}{8l}-\frac{(l^2+4m)\rho}{8l^3}
+\frac{(l^2+4m)^2\rho^2}{32l^7}+{\cal{O}}(\rho^3)\right]w_{aa}\, ,\nonumber\\
\kappa^2 <T^i_{\phantom{i}i}>&~=~&-\frac{3(l^2+4m)\rho^2}{8l^7}
-\frac{3(l^4-16m^2)\rho^3}{32l^{11}}-
\frac{3(l^4-16m^2)\rho^4}{64l^{13}}-{\cal{O}}(\rho^5)\, .
\end{eqnarray}
The leading term of the exact boundary stress tensor agrees with boundary stress tensor
calculated in the holographic renormalization method, \eq{lbs}.
The same result was founded by using an usual $r$-coordinate
\cite{Balasubramanian:1999re}. We may write the trace
of the stress tensor in terms of the temperature of the black hole and the
energy scale, as in the Poincare case. But in the case of the global patch,
the resulting expression is very complicated and we omit it here. At the UV
fixed point $\r = 0$, there is no trace anomaly as expected.
Unlike the Poincare case, the trace anomaly defined on the boundary of the pure
global AdS space-time ($m=0$), has the $\r$-dependence which is naturally coming
in to describes the volume effect of the boundary space. In other words, the conformal
symmetry near the UV fixed point is broken at the low energy scale because at this low
energy scale the volume of the boundary space becomes finite. So this finite size effect
break the conformal symmetry. For $m \ne 0$, the conformal symmetry is broken by the
thermal effect as well as the finite size effect. Interestingly, there exists a point
$\r_f = 4 l^4/(l^2 + 4m)$ where the finite size effect cancels the thermal effect
so that the trace anomaly becomes zero at this point.
\begin{figure} \label{gtra}
\begin{center}
\includegraphics[height=.3\textheight, width=.5\textwidth]{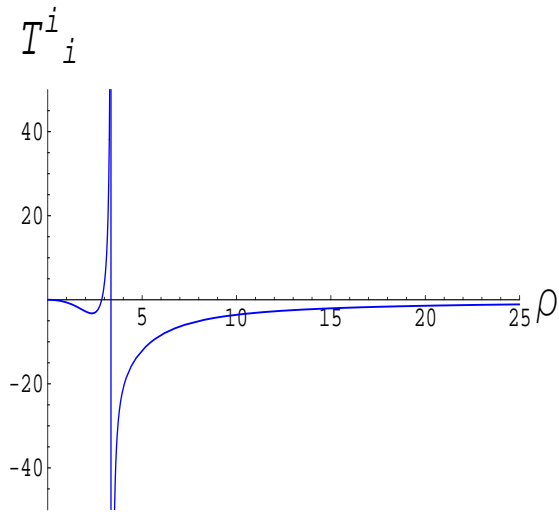}
\caption{Trace Anomaly in the Global Patch.  For convenience, we set $l=1$ and
$\kappa^2=1$. With $m=0.1$, the trace anomaly is zero at $\rho_f=2.85714$ and
diverges at the horizon $\rho_h=3.38062$.}
\end{center}
\end{figure}

As previously mentioned in the introduction, in this paper we did not consider
the logarithmic term of the boundary metric in the FG expansion and the
ignorance of this term was justified by finding the exact solution in the
FG coordinate.
Following the references \cite{deHaro:2000xn, Henningson:1998ey},
when considering the most general FG expansion including the logarithmic term,
the metric variation of the trace anomaly at the UV fixed point ($\r \to 0$)
comes from the coefficient of the logarithmic term  in the FG expansion
and the on-shell action
with the trace anomaly picks up a logarithmic divergence \cite{Emparan:1999pm}
\begin{equation}
S_{on-shell} \approx \frac{l^3}{4\kappa^2}\int d^4x
\left(\Tr[g^{-1}_{(0)}g_{(2)}g^{-1}_{(0)}g_{(2)}]
 -\Tr[g^{-1}_{(0)}g_{(2)}]\Tr[g^{-1}_{(0)}g_{(2)}]\right)
\ln\rho_0\, .
\end{equation}
In this case, using the FG expansion in \eq{dss}, the coefficient of the
logarithmic divergence term in the on-shell action is exactly zero, which
is consistent with our exact solution for the black hole solution in Poicare
and global patch.
This is also consistent with the result of other paper \cite{Burgess:1999vb}
where it was found that the ${\cal{N}}=4$ SYM theory on the product space
has no conformal anomaly.


\section{Discussion}

In this paper, we found the exact SAdS black hole solution in the
FG coordinate. Although this describes the same well-known SAdS black
hole solution, the exact one in FG coordinate gives much information for
the boundary field theory. In the usual $r$-coordinate
finding the induced metric on the boundary located at the finite distance is ambiguous
because the metric component in the radial direction depends on the boundary
position. In the FG coordinate, there is no such kind of problem since
the metric component in the radial direction does not depend on the
boundary position up to overall scale. Therefore, in the usual $r$-coordinate
the induced boundary metric is well-defined only when the boundary is
located at the infinity. To describe the Wilsonian RG of the boundary
theory we need the exact solution in the FG coordinate. Here, using
the exact black hole solutions in the FG coordinate
we have calculated the trace anomaly of the boundary theory.

In the Poincare
patch, since the boundary topology is $\textbf{R} \times \textbf{R}^3$ in the
Mikowskian version
(in the Euclidean version, $\textbf{S} \times \textbf{R}^3$), there is no scale except the
black hole mass $m$. So when $m=0$, the boundary theory has always the conformal
symmetry as the Wilsonian energy cut-off scale $\L$ runs. Considering
the black mass which gives rise to a scale,
at the UV fixed point $\r=0$ the boundary theory is still conformal but
except this UV fixed point the conformal anomaly appears
due to the black hole mass which can be reinterpreted as a thermal effect
of the boundary theory.

In the case of the global patch, the boundary topology has $\textbf{R} \times \textbf{S}^3$
in the Minkowski space-time (in the Euclidean version, the topology is
$\textbf{S} \times \textbf{S}^3$).
After the scaling of the metric by $z^2$, the boundary topology
of the spatial directions is given by $\textbf{S}^3$ where the radial coordinate
in the FG coordinate $z=\sqrt{\r}$ plays a role of the radius of the three sphere.
When the boundary is located at a finite distance of $\r$,
unlike the Poincare case, even for the pure AdS space-time ($m=0$)
there exists a trace anomaly due to the finite size effect of this three sphere.
When considering the black hole solution in the global patch, there is
another ingredient to break the conformal symmetry which is the thermal effect
like in the Poincare case.

To obtain more insight for the duality between the gravity and gauge theory,
it would be interesting to find the higher-dimensional asymptotically AdS space
time solution in the FG coordinate and more non-trivial metric containing the
effect of the other bulk fields. Furthermore, it is very interesting problem
to evaluate the central charge depending on the temperature and the RG scale
and to investigate the relation between the trace anomaly obtained here and
these central charges. This may give us clues for understanding many interesting
problems related to the RG flow like the c-theorem
and the change of the black hole entropy by factor $3/4$ when going from
strong to weak coupling regime.

\acknowledgments
We thank to Sang-Jin Sin, Ki-Myeong Lee, Ho-Ung Yee, Mu-in Park
and Yunseok Seo for helpful discussion. This work was supported by
the Science Research Center Program of the Korea Science and Engineering
Foundation through
the Center for Quantum Spacetime(CQUeST) of Sogang University with grant
number R11 - 2005 - 021.



\begin{thebibliography}{99}


\bibitem{Maldacena:1997re}
  J.~M.~Maldacena,
  ``The large N limit of superconformal field theories and supergravity,''
  Adv.\ Theor.\ Math.\ Phys.\  {\bf 2}, 231 (1998)
  [Int.\ J.\ Theor.\ Phys.\  {\bf 38}, 1113 (1999)]
  [arXiv:hep-th/9711200].


\bibitem{Gubser:1998bc}
  S.~S.~Gubser, I.~R.~Klebanov and A.~M.~Polyakov,
  ``Gauge theory correlators from non-critical string theory,''
  Phys.\ Lett.\  B {\bf 428}, 105 (1998)
  [arXiv:hep-th/9802109].


\bibitem{Witten:1998qj}
  E.~Witten,
  ``Anti-de Sitter space and holography,''
  Adv.\ Theor.\ Math.\ Phys.\  {\bf 2}, 253 (1998)
  [arXiv:hep-th/9802150].


\bibitem{Aharony:1999ti}
  O.~Aharony, S.~S.~Gubser, J.~M.~Maldacena, H.~Ooguri and Y.~Oz,
  ``Large N field theories, string theory and gravity,''
  Phys.\ Rept.\  {\bf 323}, 183 (2000)
  [arXiv:hep-th/9905111].


\bibitem{D'Hoker:2002aw}
  E.~D'Hoker and D.~Z.~Freedman,
  ``Supersymmetric gauge theories and the AdS/CFT correspondence,''
  arXiv:hep-th/0201253.


\bibitem{'tHooft:1993gx}
  G.~'t Hooft,
  ``Dimensional reduction in quantum gravity,''
  arXiv:gr-qc/9310026.


\bibitem{Susskind:1994vu}
  L.~Susskind,
  ``The World As A Hologram,''
  J.\ Math.\ Phys.\  {\bf 36}, 6377 (1995)
  [arXiv:hep-th/9409089].


\bibitem{Brown:1992br}
  J.~D.~Brown and J.~W.~.~York, "Quasilocal energy and conserved charges derived from the gravitational action,"
  Phys.\ Rev.\  {\bf D47}, 1407 (1993).

\bibitem{Balasubramanian:1999re}
  V.~Balasubramanian and P.~Kraus, "A stress tensor for anti-de Sitter gravity,"
  Commun.\ Math.\ Phys.\  {\bf 208}, 413 (1999) [arXiv:hep-th/9902121].

  S. Nojiri and S. D. Odintsov,
  "Conformal anomaly for dilaton coupled theories from AdS / CFT correspondence,"
   Phys. Lett. {\bf B444} 92 (1998); S. Nojiri and S. D. Odintsov,
   "Conformal anomaly from dS / CFT correspondence,"
   Phys. Lett. {\bf B519} 145 (2001); S. Nojiri and S. D. Odintsov,
   "Universal features of the holographic duality:
   Conformal anomaly and brane gravity trapping from 5-D AdS black hole,"
   Int. J. Mod. Phys. {\bf A18} 2001 (2003).



\bibitem{deHaro:2000xn}
  S.~de Haro, S.~N.~Solodukhin and K.~Skenderis,
  ``Holographic reconstruction of spacetime and renormalization in the  AdS/CFT
  correspondence,''
  Commun.\ Math.\ Phys.\  {\bf 217}, 595 (2001)
  [arXiv:hep-th/0002230].

\bibitem{Henningson:1998ey}
  M.~Henningson and K.~Skenderis, "Holography and the Weyl anomaly,"
  Fortsch.\ Phys.\  {\bf 48}, 125 (2000)
  [arXiv:hep-th/9812032].

  M.~Henningson and K.~Skenderis, "The holographic Weyl anomaly,"
  JHEP {\bf 9807}, 023 (1998)
  [arXiv:hep-th/9806087].

\bibitem{FG}
C. Fefferman and C. Robin Graham, ``Conformal invariants,''
The mathematical heritage of Elie Cartan (Lyon, 1984).
Asterisque 1985, Numero Hors Serie, 95--116.

\bibitem{Graham:1999jg}
  C.~R.~Graham,
  ``Volume and area renormalizations for conformally compact Einstein
  metrics,''
  [arXiv:math.dg/9909042].


\bibitem{Graham:1999pm}
  C.~R.~Graham and E.~Witten,
  ``Conformal anomaly of submanifold observables in AdS/CFT correspondence,''
  Nucl.\ Phys.\  {\bf B546}, 52 (1999)
  [arXiv:hep-th/9901021].



\bibitem{Klebanov:2000me}
  I.~R.~Klebanov,
  ``TASI lectures: Introduction to the AdS/CFT correspondence,''
  arXiv:hep-th/0009139.


\bibitem{Susskind:1998dq}
  L.~Susskind and E.~Witten,
  ``The holographic bound in anti-de Sitter space,''
  arXiv:hep-th/9805114.


\bibitem{deBoer:1999xf}
  J.~de Boer, E.~P.~Verlinde and H.~L.~Verlinde,
  ``On the holographic renormalization group,''
  JHEP {\bf 0008}, 003 (2000)
  [arXiv:hep-th/9912012]

  E.~P.~Verlinde and H.~L.~Verlinde,
  ``RG-flow, gravity and the cosmological constant,''
  JHEP {\bf 0005}, 034 (2000)
  [arXiv:hep-th/9912018]

  J.~de Boer,
  ``The holographic renormalization group,''
  Fortsch.\ Phys.\  {\bf 49}, 339 (2001)
  [arXiv:hep-th/0101026].


\bibitem{Imbimbo:1999bj}
  C.~Imbimbo, A.~Schwimmer, S.~Theisen and S.~Yankielowicz,
  ``Diffeomorphisms and holographic anomalies,''
  Class.\ Quant.\ Grav.\  {\bf 17}, 1129 (2000)
  [arXiv:hep-th/9910267].

  A.~Schwimmer and S.~Theisen,
  ``Diffeomorphisms, anomalies and the Fefferman-Graham ambiguity,''
  JHEP {\bf 0008}, 032 (2000)
  [arXiv:hep-th/0008082].

A.~Schwimmer and S.~Theisen,
  ``Universal features of holographic anomalies,''
  JHEP {\bf 0310}, 001 (2003)
  [arXiv:hep-th/0309064].


\bibitem{Nakamura:2006ih}
  S.~Nakamura and S.~J.~Sin,
  ``A holographic dual of hydrodynamics,''
  JHEP {\bf 0609}, 020 (2006)
  [arXiv:hep-th/0607123].


\bibitem{Myers:1999ps}
  R.~C.~Myers,
  ``Stress tensors and Casimir energies in the AdS/CFT correspondence,''
  Phys.\ Rev.\  D {\bf 60}, 046002 (1999)
  [arXiv:hep-th/9903203].


\bibitem{Horowitz:1998ha}
  G.~T.~Horowitz and R.~C.~Myers,
  ``The AdS/CFT correspondence and a new positive energy conjecture for
  general relativity,''
  Phys.\ Rev.\  D {\bf 59}, 026005 (1999)
  [arXiv:hep-th/9808079].


\bibitem{Gubser:1996de}
  S.~S.~Gubser, I.~R.~Klebanov and A.~W.~Peet,
  ``Entropy and Temperature of Black 3-Branes,''
  Phys.\ Rev.\  D {\bf 54}, 3915 (1996)
  [arXiv:hep-th/9602135].



\bibitem{Burgess:1999vb}
  C.~P.~Burgess, N.~R.~Constable and R.~C.~Myers,
  ``The free energy of N = 4 superYang-Mills and the AdS/CFT  correspondence,''
  JHEP {\bf 9908}, 017 (1999)
  [arXiv:hep-th/9907188].


\bibitem{Emparan:1999pm}
  R.~Emparan, C.~V.~Johnson and R.~C.~Myers,
  Phys.\ Rev.\  D {\bf 60}, 104001 (1999)
  [arXiv:hep-th/9903238].




\end{thebibliography}
\end{document}